\documentclass[twocolumn,pra,showpacs]{revtex4}
\usepackage{amsmath}
\usepackage{graphicx}
\newcommand{\ket}[1]{\ensuremath{|#1\rangle}}
\newcommand{\bra}[1]{\ensuremath{\langle#1|}}
\newcommand{\nuc}[2]{\mbox{${}^{#1}\rm #2$}}

\begin{document}
\title{Error tolerance in an NMR Implementation of \\ Grover's Fixed-Point Quantum Search Algorithm}
\author{Li Xiao}
\author{Jonathan A. Jones}\email{jonathan.jones@qubit.org}
\affiliation{Centre for Quantum Computation, Clarendon Laboratory,
University of Oxford, Parks Road, OX1 3PU, United Kingdom}
\date{\today}
\begin{abstract}
We describe an implementation of Grover's fixed-point quantum search
algorithm on a nuclear magnetic resonance (NMR) quantum computer,
searching for either one or two matching items in an unsorted
database of four items.  In this new algorithm the target state (an
equally weighted superposition of the matching states) is a fixed
point of the recursive search operator, and so the algorithm always
moves towards the desired state.  The effects of systematic errors
in the implementation are briefly explored.
\end{abstract}
\maketitle

\section{Introduction}
Grover's quantum search \cite{grover97, grover98} is one of the key
algorithms in quantum computation \cite{bennett00}, allowing an
unstructured database to be searched more efficiently than can be
achieved by any classical algorithm.  It is most simply described in
terms of a binary function $f$, with an $n$ bit input register,
permitting $N=2^n$ inputs, and a single output bit. This function
has the value $1$ for $k$ desired (\textit{matching} or
\textit{satisfying}) inputs and zero for the other $N-k$ inputs, and
can only be investigated by an oracle which returns the value for
any desired input. Grover's original algorithm uses a quantum oracle
which performs the transformation
\begin{equation}
\ket{x}\longrightarrow(-1)^{f(x)}\ket{x}
\end{equation}
which is applied alternately with an amplitude amplification
operator.  Beginning from the state $H_n\ket{\mathbf{0}}$, where
$H_n$ is the $n$ qubit Hadamard and \ket{\mathbf{j}} indicates an
$n$ qubit quantum register containing the number $j$, the system
is rotated towards an equally weighted superposition of the
satisfying inputs.  A measurement of the register after
$O(\sqrt{N/k})$ steps will return one of the satisfying inputs
with high probability, while a classical search will take $O(N/k)$
queries on average.

Grover's search is known to be optimal \cite{zalka99} when the
number of matching inputs is known, but problems occur when $k$ is
unknown. As the basic procedure is a rotation, once the desired
state is reached further iterations will drive the system
\textit{away} from this state. Thus it is necessary either to
estimate the value of $k$ (for example, by approximate quantum
counting \cite{boyer98, jones99}) or to use an algorithm which is
more robust to errors in the value of $k$.

Recently Grover has described a new quantum algorithm
\cite{newgrover}, which overcomes this problem by driving the system
asymptotically towards the desired state: the algorithm will always
move towards the target and cannot overshoot.  This might seem
impossible, as unitarity means that any iterative algorithm cannot
have a fixed point, but the new algorithm overcomes this by using a
process which is recursive rather than iterative, and so the target
state can act as a fixed point.  A further consequence of the
fixed-point behavior is that the new algorithm should be relatively
robust to certain types of systematic error in its implementation.

\section{Theory}
Grover's algorithm comes in many forms \cite{grover98}, and here we
describe just one of these, before relating it to the new algorithm
\cite{newgrover}. Consider the transformation
\begin{equation}
UR_0U^\dag R_fU\ket{\mathbf{0}}
\end{equation}
where $R_f$ is a phase oracle which applies a phase of $\phi$ to all
basis states satisfying the function $f$ and $R_0$ applies this
phase shift to the initial state \ket{\mathbf{0}}.  If we take
$U=H_n$ and $\phi=\pi$ then this corresponds to the first iteration
of the original Grover algorithm. Subsequent steps are obtained by
applying the last four operations $r$ times, thus applying
successive rotations. Larger rotations can be defined using a
recursive approach, by taking
\begin{equation}
V_{r+1}=V_rR_0V^\dag_rR_fV_r
\end{equation}
with $V_0=U$.  For the original search algorithm $U^\dag=U$ and
$R^\dag=R$, and so each recursive operator simply corresponds to one
of the iterative operators.

Grover's new fixed point quantum search differs significantly from
this by choosing $\phi=\pi/3$, so that $R^\dag$ does not equal $R$.
Thus the recursive operators are not simply iterative, and have to
be worked out separately for each value of $r$; for more details see
the original paper \cite{newgrover}.  Here we consider the case of
$n=2$, with either $k=1$ or $k=2$, and take $U$ as the
pseudo-Hadamard gate (a $90^\circ_y$ rotation), as these are the
examples we implement experimentally.

For the case of $k=1$ there is a single satisfying input \ket{s}.
The probability of the algorithm succeeding depends on the
projection of the final state onto the satisfying input and is given
by
\begin{equation}
P_r=|\bra{s}V_r\ket{00}|^2 \label{eq:Pr1}
\end{equation}
which simplifies to
\begin{equation}
P_r=1-(3/4)^{3^r}.
\end{equation}
Clearly this converges rapidly to one, as shown by the numerical
values listed in table~\ref{tab:Pr}.  For the case $k=2$ the target
state is an equally weighted superposition of the two satisfying
states, and the success probability is given by
\begin{equation}
P_r=1-(1/2)^{3^r} \label{eq:Pr3}
\end{equation}
which rises even more rapidly.

\begin{table}
\caption{Success probabilities, $P_r$, and the total number of
queries used, $Q_r$, for the $r^\text{th}$ stage of Grover's
fixed-point quantum search algorithm with $n=2$; for more details
see the main text.} \label{tab:Pr}
\begin{ruledtabular}
\begin{tabular}{llll}
$r$&$P_r\;(k=1)$&$P_r\;(k=2)$&$Q_r$\\\hline
0&0.2500&0.5000&0\\
1&0.5781&0.8750&1\\
2&0.9249&0.9980&4\\
3&0.9996&1.0000&13\\
4&1.0000&1.0000&40
\end{tabular}
\end{ruledtabular}
\end{table}

Note, however, that $r$ is the order of the recursive operator and
not the number of queries, which is given by $Q_r=(3^r-1)/2$.  This
query count, which is also listed in Table~\ref{tab:Pr}, also rises
rapidly with $r$, so the fixed-point algorithm is less efficient
than a traditional Grover search when the value of $k$ is known.
This is unsurprising, as the traditional search is known to be
optimal \cite{zalka99} in this case!  When the value of $k$ is
unknown, however, the new algorithm outperforms previously known
methods, as described by Grover \cite{newgrover}.  It is also
relatively robust against experimental errors, as discussed below.

\section{Experiment}
This algorithm was implemented on a two qubit nuclear magnetic
resonance (NMR) quantum computer \cite{cory96, cory97,
gershenfeld97, jones98c, jones01a, vandersypen04}, searching for
either one or two satisfying inputs for a function with four inputs.
The spin system chosen was formed by the \nuc{1}{H} and \nuc{13}{C}
nuclei in a sample of 10\,mg of \nuc{13}{C} labeled sodium formate
($\text{Na}^+\text{HCO}_2^-$) dissolved in 0.75\,ml of
$\text{D}_2\text{O}$ at a temperature of $20^\circ\text{C}$.

All experiments were performed on a Varian Unity Inova spectrometer
with a nominal \nuc{1}{H} frequency of 600\,MHz. The \nuc{1}{H} and
\nuc{13}{C} frequencies were adjusted to be in exact resonance with
the respective nuclei so that the spin Hamiltonian in the rotating
frame, written using Product Operator notation \cite{sorensen83}, is
given by the Ising coupling
\begin{equation}
\mathcal{H}=\pi J\,2H_zC_z
\end{equation}
with $J=194.8\,\text{Hz}$.  The measured relaxation times were
$\text{T}_1^\text{H}=6.5\,\text{s}$,
$\text{T}_2^\text{H}=1.2\,\text{s}$,
$\text{T}_1^\text{C}=16\,\text{s}$, and
$\text{T}_2^\text{C}=0.6\,\text{s}$.  A repetition delay of 120\,s
was used in all experiments; this more than seven times the longest
$\text{T}_1$ and so saturation effects can be ignored. The radio
frequency (rf) pulse powers were adjusted so that a $90^\circ$
rotation took $15\,\mu\text{s}$ for both spins.

The new search algorithm requires the implementation of $U$ gates
and $R$ gates, as well as the inverse operations.  The $U$ gates
were implemented as simultaneous $90^\circ_y$ pulses, while the $R$
gates were decomposed into periods of evolution under the Ising
coupling and composite $z$-rotations \cite{freeman97b}, constructed
from $x$ and $y$-pulses.  $R_f$ gates were implemented for each of
the four functions $f$ with $k=1$ and the six functions with $k=2$;
note that the $R_0$ gate is identical to $R_f$ for the function with
\ket{00} as the single satisfying input. Each gate was locally
optimized by combining pulses, but no optimization across gates was
performed. To explore the effects of systematic errors, experiments
were performed using both naive rf pulses and BB1 composite pulses
\cite{wimperis94, cummins02} which correct for systematic errors in
pulse lengths arising from rf inhomogeneity.

A pseudo-pure initial \ket{00} state was prepared by spatial
averaging \cite{jones01a}. Experiments were performed for each of
the functions with $k=1$ and $k=2$, with the order $r$ of the
recursive search operator taking the values $r=0,\,1,\,2,\,3$.  The
$r\rightarrow\infty$ limit was simulated by directly transforming
the initial state into the desired target state. The state of the
spin system was then probed to obtain information on the performance
of the algorithm.

For the case of $k=1$ the target state is a single eigenstate and
the analysis is simple. A crush gradient was applied to dephase any
off-diagonal error terms in the density matrix \cite{jones98b} and
the \nuc{1}{H} NMR spectrum was observed after a $90^\circ_y$
\nuc{1}{H} pulse. The state of the first qubit, stored on the
\nuc{1}{H} nucleus, is then encoded in the sign of the NMR
resonance, and the state of the second qubit, stored on the
\nuc{13}{C} nucleus, is revealed by which of the two components of
the \nuc{1}{H} multiplet is observed.  For more details see
\cite{jones01a}.  Note that this approach is only practical in
systems with relatively small numbers of qubits, and in large spin
systems it would be necessary to measure all the spin states
directly.

Finally, the success probability $P_r$ of the algorithm can be
estimated from the intensity of the NMR spectrum compared with a
reference spectrum \cite{anwar05a}. Because NMR experiments are only
sensitive to the traceless part of the density matrix the observed
signal strength is not directly proportional to $P_r$, but rather to
the fractional signal $F_r$, which is given by $F_r=(4P_r-1)/3$. In
particular, no signal is expected for the case $P_r=1/4$.

\begin{figure*}
\includegraphics{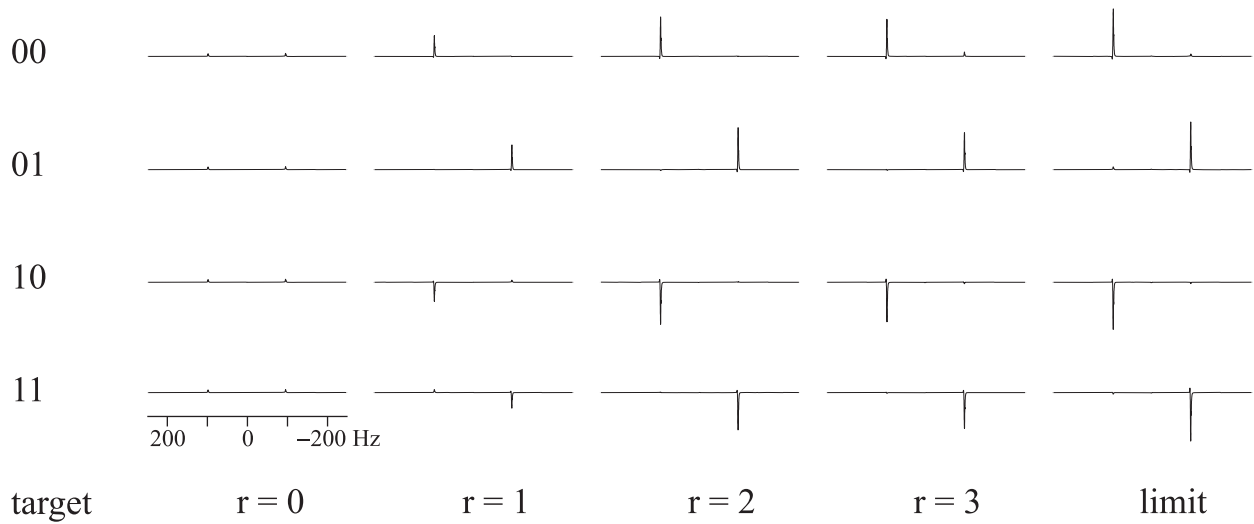}
\caption{Experimental \nuc{1}{H} NMR spectra from an implementation
of Grover's fixed-point quantum search algorithm on a two qubit NMR
quantum computer with one matching state.  For a description of the
readout scheme see the main text.  Spectra are shown for the cases
$r=0,\,1,\,2,\,3$, and a simulation of the $r\rightarrow\infty$
limit was obtained by directly transforming the initial state into
the target state. Spectra are plotted using NMR conventions, with
frequency (measured by the offset from the rf frequency) increasing
from right to left. A horizontal axis is plotted below the bottom
left spectrum and can be applied to all spectra.  The vertical scale
is arbitrary, but the same for all spectra.} \label{fig:one}
\end{figure*}

For the case of $k=2$ the situation is slightly more complicated, as
the target state is a superposition of the two matching states, but
we chose to analyze the data in the same way.  For four of the six
functions this results in a \nuc{1}{H} NMR spectrum containing
\textit{both} components of the multiplet, with the result encoded
in the signs of these two lines, while for the other two functions
no signal is expected.  For the four functions giving rise to
visible signals the success probability and fractional signal are
related by $F_r=2P_r-1$, and so no signal is expected for the case
$P_r=1/2$.

\section{Results}
We began by implementing the four functions with $k=1$ using naive
rf pulses, with the results shown in Fig.~\ref{fig:one}. The spectra
all have the form expected, showing one major component in each
multiplet.  A positive line indicates that the first qubit is in
state \ket{0} while a negative line indicates state \ket{1}. Signal
in the left hand component indicates that the second qubit is in
state \ket{0} while the right hand component indicates state
\ket{1}.  The minor signals visible on the other component of each
multiplet, as well as the minor phase distortions visible in some
spectra, can be ascribed to errors in the implementation.

\begin{figure*}
\includegraphics{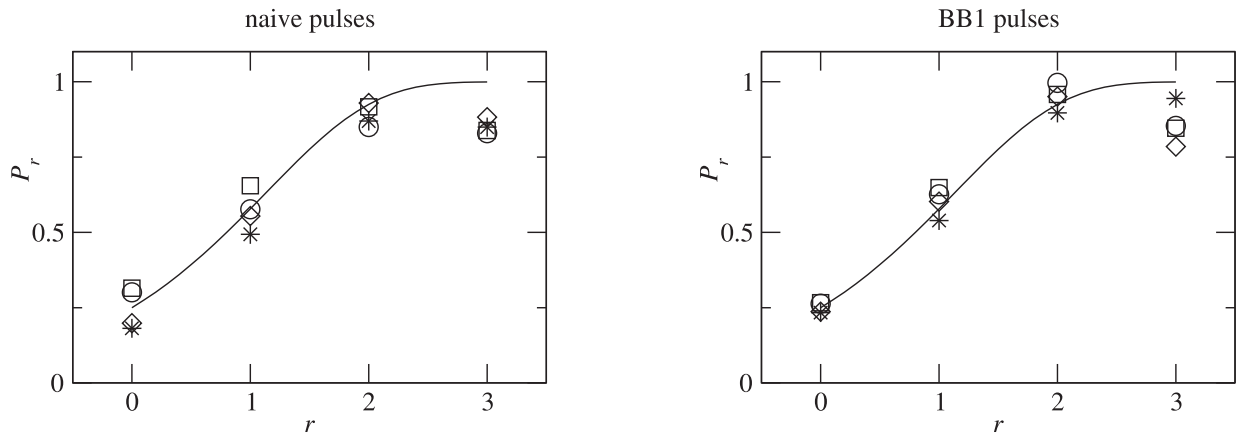}
\caption{Experimental success probability for the cases
$r=0,\,1,\,2,\,3$ for the four possible functions with a single
matching state.  Results are shown for both simple single qubit
gates implemented by naive rf pulses and for BB1 composite pulses.
Fractional intensities are obtained by comparing the spectral
intensities with those in the spectra obtained by directly
transforming the initial state into the target state, and these are
converted to probabilities as described in the main text. The
theoretical result is shown as a smooth curve even though it is
strictly only defined for integer values of $r$; experimental values
are shown as squares, circles, diamonds and stars for target states
of \ket{00}, \ket{01}, \ket{10} and \ket{11} respectively.}
\label{fig:two}
\end{figure*}

As expected the signal intensity initially rises towards the
limiting value, although it seems to fall slightly at $r=3$.  This
point is explored in more detail in Fig.~\ref{fig:two} which
compares the integrated intensity of the largest component with the
theoretically expected values. Initially the experimental data
points lie quite close to the theoretical line, but for $r=3$ the
match is much less good. We originally ascribed this to the effects
of errors in the pulse sequence, and in particular to the cumulative
effects of pulse length errors, and sought to reduce these effects
by using BB1 composite pulses \cite{wimperis94, cummins02}.  The
outline results of this approach are also shown in
Fig.~\ref{fig:two} (raw data not shown).

While this approach did give slightly improved results for $r\le2$
(especially for the case $r=0$ where the search operator comprises a
single $90^\circ$ pulse), it does not remove the drop in intensity
seen at $r=3$, which we now believe arises from errors in the
implementation of the $R$ gates. The errors are different for the
four different functions (the error arising from noise in the
experimental spectra was estimated by repetition and is much smaller
than the scatter observed), consistent with this suggestion.

The lack of improvement from the use of BB1 pulses may seem
disappointing, but is in fact quite interesting in its own right. We
have assumed that the $U$ operator is implemented by a $90^\circ_y$
pulse, but Grover's algorithm can be made to work with many
different operators \cite{grover98}.  In combination with the fact
that Grover's new algorithm always moves towards the target state,
this makes the algorithm intrinsically tolerant of pulse length
errors \cite{newgrover}.  In fact the experimental spectra observed
are of remarkably high quality, given that the case of $r=3$
corresponds to the implementation of 26 two qubit gates (13
instances of $R_f$ and 13 of $R_0$) and around 200 rf pulses.

Finally we consider the situation when there are two matching
states, that is $k=2$.  The search operators were implemented
directly, rather than by applying two single-match operators in
sequence, and are slightly simpler than for the case of a single
target state (in some cases the function operators $R_f$ do not
require two qubit gates). There are six possible search operators
$R_f$, all of which were implemented using both naive and BB1
composite pulses, but here we concentrate on two cases: firstly
where the target states are \ket{00} and \ket{01} (giving a
\nuc{1}{H} spectrum with both components of the multiplet positive),
and secondly where they are \ket{10} and \ket{01} (giving a spectrum
with the lefthand component negative and the right hand component
positive).

\begin{figure}
\includegraphics{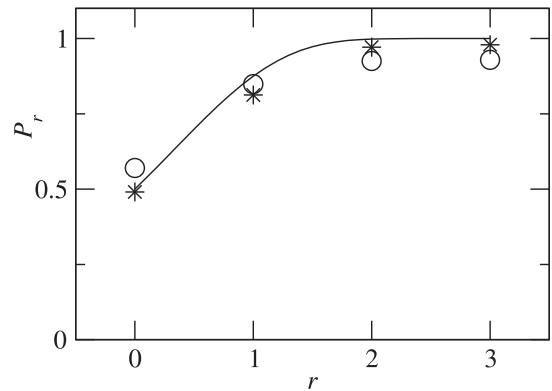}
\caption{Experimental success probability for the cases
$r=0,\,1,\,2,\,3$ for two of the six possible functions with two
matching states.  For further details see Fig.~\ref{fig:two}.
Experimental values were obtained using naive pulses and are shown
as circles for matching states of \ket{00} and \ket{01} and stars
for matching states of \ket{10} and \ket{01}.} \label{fig:three}
\end{figure}

The experimental results from these cases are summarized in
Fig.~\ref{fig:three}, where the intensity of each spectrum was
obtained using either the sum or the difference of the integrals of
the two components as appropriate. We show results obtained using
naive pulses, but as before the results with BB1 pulses were very
similar. As before the experimental data broadly follows the
theoretical curve, and this time the results for $r=2$ and $r=3$
have almost the same intensity, as predicted by Table~\ref{tab:Pr}.
This slight improvement may reflect the fact that the $R_f$ gates
are slightly simpler for $k=2$ than for $k=1$.

In principle one could also study the cases of $k=3$ and $k=4$, but
these are not particularly interesting.  There is a correspondence
between the functions with $k$ and those with $N-k$ satisfying
inputs, and for quantum oracles (controlled phase gates) the
operators differ only by global phases.  Thus the case of $k=3$ is
indistinguishable from that of $k=1$, as phase shifting three states
in one direction is equivalent to shifting the fourth state in the
opposite direction. The case of $k=4$ is trivial, as applying phase
shifts to all the states is nothing more than a global phase, and
thus the function operator corresponds to the identity operation.

\section{Conclusions}
Our experimental results are broadly consistent with those predicted
for an implementation of Grover's new fixed-point quantum search
algorithm.  The observed success probability initially rises with
the order of the (recursively defined) search operator, although a
fall-off is observed at $r=3$, which we ascribe to experimental
errors in the operators $R_f$ and $R_0$.  Despite these experimental
errors, the results are remarkably good given the complexity of the
pulse sequences involved.

As predicted by Grover \cite{newgrover}, the algorithm is remarkably
robust to systematic errors which are equivalent for a gate and its
inverse.  This is largely true of the $U$ gates, which are
implemented using rf pulses, and the use of BB1 gates to correct
systematic errors has little effect except in the case $r=0$.  It is
not true for the $R$ gates, as our implementation of $R^\dag$ is
somewhat different from that of $R$, and the errors in these two
gates will not be equivalent.

This error-tolerance property can in principle be used to develop
methods for more general correction of systematic errors
\cite{newgrover}, but we do not address this point here.

\begin{acknowledgments}
We thank the UK EPSRC and BBSRC for financial support.
\end{acknowledgments}


\begin{thebibliography}{99}
\bibitem{grover97} L.~K. Grover, Phys. Rev. Lett. \textbf{79,} 325
(1997).

\bibitem{grover98} L.~K. Grover, Phys. Rev. Lett. \textbf{80,} 4329 (1998).

\bibitem{bennett00}
C.~H. Bennett and D.~P. DiVincenzo, Nature (London) \textbf{404,}
247 (2000).

\bibitem{zalka99}
C.~Zalka, Phys. Rev. A \textbf{60,} 2746 (1999).

\bibitem{boyer98}
M.~Boyer, G.~Brassard, P.~H{\o}yer and A.~Tapp, Fort. Der. Physik
\textbf{46,} 493 (1998).

\bibitem{jones99}
J.~A. Jones and M.~Mosca, Phys. Rev. Lett. \textbf{83,} 1050 (1999).

\bibitem{newgrover}
L.~K. Grover, \textit{A different kind of quantum search},
\eprint{quant-ph/0503205}.

\bibitem{cory96}
D.~G. Cory, A.~F. Fahmy and T.~F. Havel, in \textit{Proceedings of
PhysComp '96} (New England Complex Systems Institute, Cambridge MA,
1996).

\bibitem{cory97}
D.~G. Cory, A.~F. Fahmy and T.~F. Havel, Proc. Nat. Acad. Sci. USA
\textbf{94,} 1634 (1997).

\bibitem{gershenfeld97}
N.~A. Gershenfeld and I.~L. Chuang, Science \textbf{275,} 350
(1997).

\bibitem{jones98c}
J.~A. Jones and M.~Mosca, J. Chem. Phys. \textbf{109,} 1648.

\bibitem{jones01a}
J.~A. Jones, Prog. NMR. Spectrosc. \textbf{38,} 325 (2001).

\bibitem{vandersypen04}
L.~M.~K. Vanderspen and I.~L. Chuang, Rev. Mod. Phys. \textbf{76,}
1037 (2004).

\bibitem{sorensen83}
O.~W. S{\o}rensen, G.~W. Eich, M.~H. Levitt, G.~Bodenhausen and
R.~R. Ernst, Prog. NMR. Spectrosc. \textbf{16,} 163 (1983).

\bibitem{freeman97b}
R.~Freeman, \textit{Spin Choreography} (Spektrum, Oxford, 1997).

\bibitem{wimperis94}
S.~Wimperis, J. Magn. Reson. A \textbf{109,} 221 (1994).

\bibitem{cummins02}
H.~K. Cummins, G.~Llewellyn and J.~A. Jones, Phys. Rev. A
\textbf{67,} 042308 (2003).

\bibitem{jones98b}
J.~A. Jones, M.~Mosca and R.~H. Hansen, Nature \textbf{393,} 344
(1998).

\bibitem{anwar05a}
M.~S. Anwar, L.~Xiao, A.~J. Short, J.~A. Jones, D.~Blazina, S.~B.
Duckett, and H.~A. Carteret, Phys. Rev. A \textbf{71,} 032327
(2005).


\end{thebibliography}
\end{document}